\RequirePackage{fix-cm}
\documentclass[twocolumn]{svjour3}          % twocolumn
\smartqed  % flush right qed marks, e.g. at end of proof
\usepackage{graphicx}
\usepackage{mathptmx}      % use Times fonts if available on your TeX system
\usepackage{latexsym}

\usepackage[T1]{fontenc}
\usepackage[utf8]{luainputenc}
\usepackage{amsmath}
\usepackage{amssymb}
\usepackage{wasysym}
\usepackage{microtype}
\usepackage{color}
\usepackage[misc]{ifsym}
\usepackage[square, comma, sort&compress, numbers]{natbib}
\usepackage[numbers]{natbib}
\begin{document}

\title{Narrow-linewidth cooling of $^{6}$Li atoms using the 2S-3P transition%\thanks{Grants or other notes
}


\author{Hao-Ze Chen$^{1,2,3}$\and Xing-Can Yao$^{1,2,3,4}$\and Yu-Ping Wu$^{1,2,3}$\and Xiang-Pei Liu$^{1,2,3}$\and Xiao-Qiong Wang$^{1,2,3}$\and Yu-Ao Chen$^{1,2,3}$\and Jian-Wei Pan$^{1,2,3,4}$}


\institute{
%\begin{table*}
%\begin{tabular}{ll}
%\Letter & Jian-Wei Pan
%\end{tabular}
%\end{table*}
\begin{enumerate}
  \item [$^1$] Shanghai Branch, National Laboratory for Physical Sciences at Microscale and Department of Modern Physics, University of Science and Technology of China, Hefei, Anhui 230026, China
  \item [$^2$] CAS Center for Excellence and Synergetic Innovation Center in Quantum Information and Quantum Physics, University of Science and Technology of China, Shanghai, 201315, China
  \item [$^3$] CAS-Alibaba Quantum Computing Laboratory, Shanghai, 201315, China
  \item [$^4$] Physikalisches Institut, Ruprecht-Karls-Universität Heidelberg, Im Neuenheimer Feld 226, 69120 Heidelberg, Germany
\end{enumerate}
}

\date{Received: date / Accepted: date}

\maketitle

\begin{abstract}

We report on a narrow-linewidth cooling of $^{6}$Li atoms using the $2S_{1/2}\to 3P_{3/2}$ transition in the ultraviolet (UV) wavelength regime. By combining the traditional red magneto-optical trap (MOT) at 671 nm and the UV MOT at 323 nm, we obtain a cold sample of $1.3\times10^9$ atoms with a temperature of 58 $\mu$K. Furthermore, we demonstrate a high efficiency magnetic transport for $^{6}$Li atoms with the help of the UV MOT. Finally, we obtain $8.1\times10^8$ atoms with a temperature of 296 $\mu$K at a magnetic gradient of 198 G/cm in the science chamber with a good vacuum environment and large optical access.

\end{abstract}

\section{Introduction}

The remarkable achievements of ultracold quantum gases not only open the way for high-precision quantum metrology~\cite{peters2001high}, but also shed new light on the study of few or many-body quantum physics~\cite{bloch2008many,bloch2012quantum}. Today, a total of thirteen elements are available in the ultracold quantum gas experiments due to the dramatic advances in the laser technology and more sophisticated cooling techniques. Among them, dilute Fermi gases of $^{6}$Li atoms have attracted intense interests both in experimental and theoretical studies, due to its broad Feshbach resonance at 832 G~\cite{zurn2013precise}. It offers a great opportunity to study both weak-coupling Bardeen-Cooper-Schrieffer (BCS) superfluid and Bose-Einstein condensation (BEC) of tightly bound fermion pairs~\cite{chen2005bcs}. Several breakthroughs have been demonstrated with dilute gases of $^{6}$Li, such as the realization of a molecular BEC~\cite{jochim2003bose,zwierlein2003observation}, the observation of pairing and phase separation in a polarized Fermi gas~\cite{partridge2006pairing,zwierlein2006direct}, and the observation of vortex lattices in a Fermi gas~\cite{zwierlein2005vortices}, etc.

Although great successes have been achieved in such a system, the production of large quantum degenerate gases of $^{6}$Li is still challenging without sympathetic cooling~\cite{hadzibabic2003fiftyfold}. For most alkali-metal atoms, Sub-Doppler temperatures can be attained via the Sisyphus cooling mechanism in an optical molasses~\cite{ungar1989optical,lett1989optical}. However, this mechanism fails when the natural linewidth is comparable to the excited hyperfine splitting~\cite{xu2003single}. Many elements, e.g. lithium and potassium, suffer from the unresolved D2 excited hyperfine structures that inhibit the conventional sub-Doppler cooling process. Tremendous efforts have been devoted to achieving temperatures lower than the D2 Doppler cooling limit for these atomic gases. The near detuned sub-Doppler cooling has been successfully applied for the bosonic potassium isotopes using properly selected detunings and intensities~\cite{landini2011sub}. Recently, a so-called gray molasses sub-Doppler cooling technique that takes advantages of both the D1-line Sisyphus cooling and velocity selective coherent population trapping (VSCPT), has been demonstrated for lithium and potassium atoms~\cite{sievers2015simultaneous,burchianti2014efficient,grier2013lambda,nath2013quantum,salomon2014gray,fernandes2012sub,chen2016production}.

Laser cooling on narrow optical transitions is another approach that is widely used in alkaline-earth metal systems to achieve lower temperatures~\cite{katori1999magneto,binnewies2001doppler}. This scheme has also been demonstrated in the alkali metal atoms for transitions from the ground state $nS_{1/2}$ to the excited state $(n+1)P_{3/2}$~\cite{duarte2011all,sebastian2014two,mckay2011low}. The linewidths of these transitions are much narrower than those of the D2 transitions, resulting in a much lower Doppler temperature. For instance, the overall linewidth of the $2S_{1/2}\to 3P_{3/2}$ transition of $^{6}$Li at 323 nm is $\Gamma_{3P}=2\pi\times754$ kHz, corresponding to a Doppler temperature of 18 $\mu$K that is much lower than the D2-line Doppler temperature of 141 $\mu$K. Temperatures as low as 59 $\mu$K and 33 $\mu$K have been experimentally achieved for $^{6}$Li atoms in UV MOTs using this transition~\cite{duarte2011all,sebastian2014two}.

Here, we report on a narrow-linewidth cooling of $^{6}$Li atoms using the $2S_{1/2}\to 3P_{3/2}$ transition at 323 nm. It allows us to lower the initial temperature of $1.3\times10^{9}$ atoms from 252 $\mu$K to 58 $\mu$K. Moreover, we demonstrate the advantage of this laser cooling technique by transporting the magnetically confined atoms from the MOT chamber to the science chamber. We show that, with the help of the UV MOT, the magnetic transport efficiency increases from 23.6\% to 70.3\%, which  greatly benefits the subsequent evaporative cooling. Our results pave the way for the production of large degenerate Fermi gases of $^{6}$Li atoms.

\section{Experimental setup}
\label{sec:1}

323 nm laser source plays an essential role in narrow-linewidth cooling of $^{6}$Li atoms. It can be generated via second-harmonic generation (SHG) with a 646 nm laser. Laser power up to 200 mW can be obtained using tapered amplifier, where the linewidth is around 1 MHz at 646 nm. Another approach utilizes sum frequency generation (SFG) of two infrared fiber lasers (e.g., 1100 nm and 1570 nm) through a periodically poled lithium niobate (PPLN) crystal~\cite{wilson2011750}. Although both high power ($\sim$ 1 Watt) and narrow-linewidth ($\sim$ 100 kHz) 646 nm laser can be obtained using this method, it also increases the experimental cost and complexity. Therefore, we adopt a commercial semiconductor SHG laser system (Toptica SHG-Pro). With a 200 mW fundamental input at 646 nm, the maximum second harmonic output is 60 mW.

\begin{figure}[htbp]
\centering
\includegraphics[width=\columnwidth]{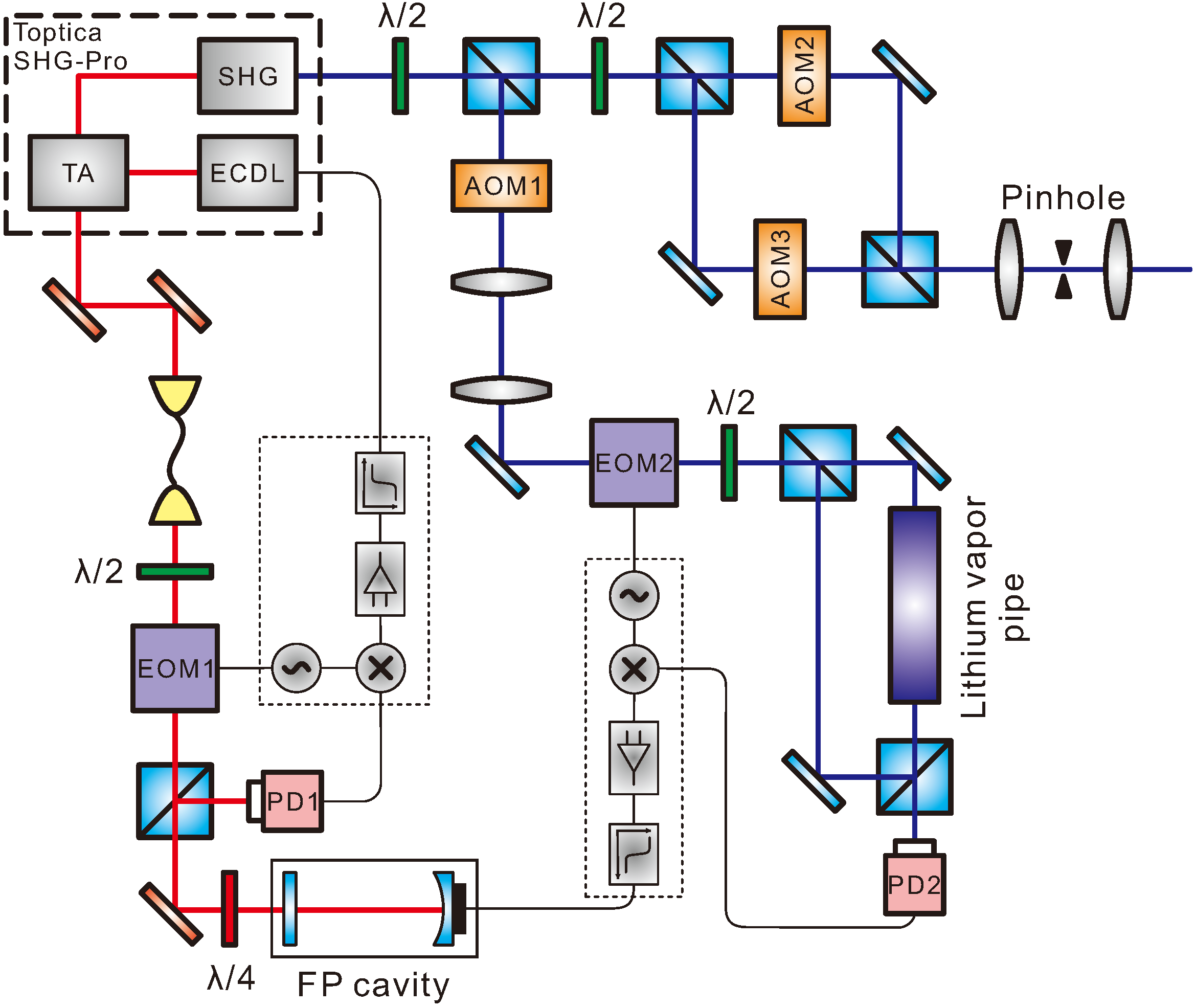}
\caption{(Color online). Optical setup for the UV MOT at 323 nm. The laser source (Toptica SHG-Pro) consists of an external cavity diode laser, a tapered amplifier and a frequency doubler. In order to narrow the linewidth, the fundamental laser at 646nm is locked to a homemade FP cavity through the PDH method. The cavity is further locked to the $^{6}$Li UV transition to ensure the long-term frequency stabilization. The main output of the UV laser is divided into two parts. They pass through two AOMs for shifting the frequency of -114 MHz and +114 MHz, serving as the cooling and repumping lasers for the UV MOT, respectively. Then, the two laser beams are superimposed on a PBS and pass through a spatial filter to generate a TM00 mode beam.}
\label{Fig1}
\end{figure}

\begin{figure}[htbp]
\centering
\includegraphics[width=0.8\columnwidth]{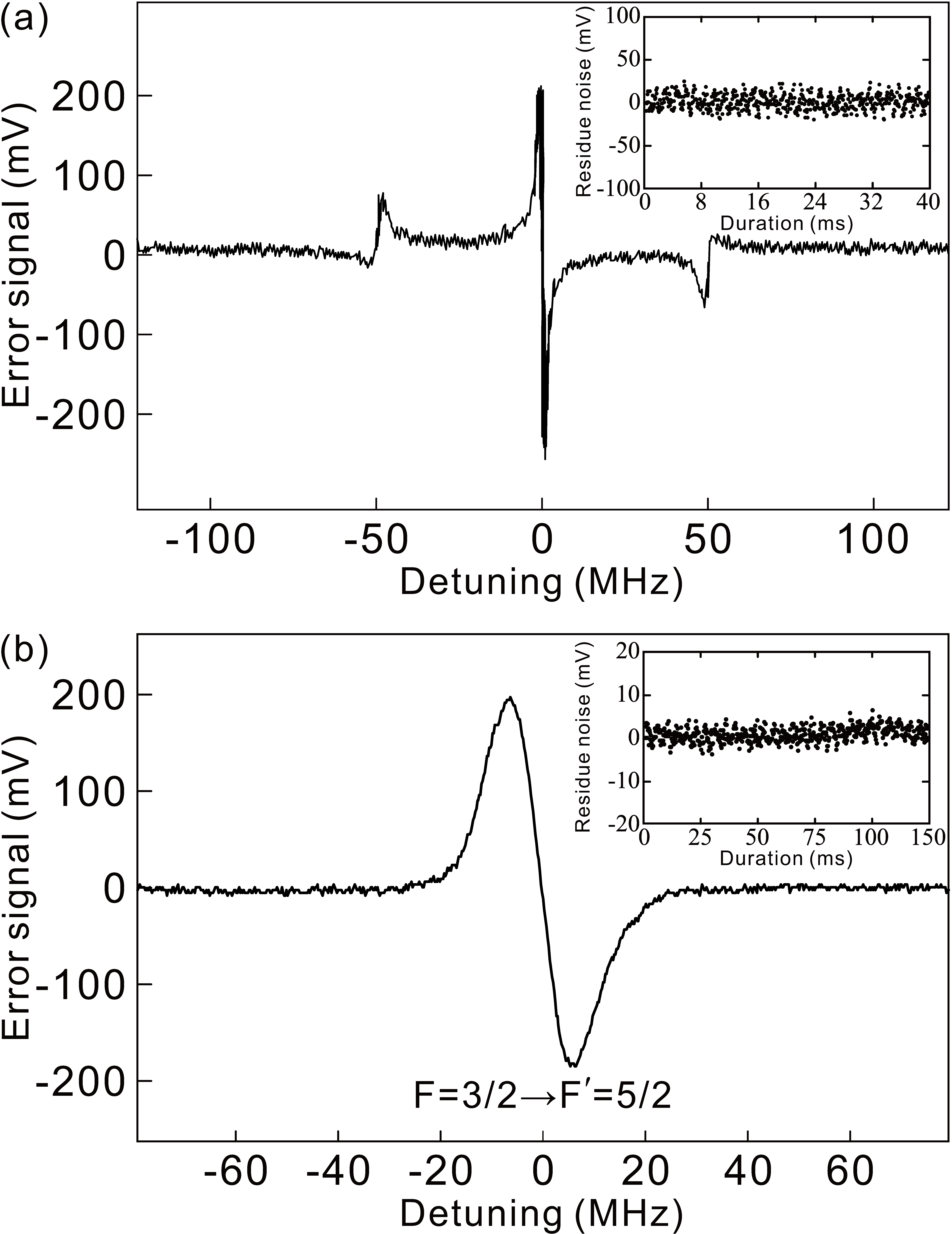}
\caption{(a) and (b) show the error signal of the FP cavity and $|2S_{1/2},\ F=3/2\rangle\to |3P_{3/2},\ F'=5/2\rangle$ transition of $^{6}$Li, respectively. The insets show the residual noises after frequency stabilization. After the two-stage frequency locking, the obtained linewidth of the UV laser is about 68 kHz on the order of tens of milliseconds.}
\label{Fig2}
\end{figure}

Our optical setup for the UV MOT is shown in Fig.~\ref{Fig1}. To achieve a UV MOT, linewidth reducing and frequency locking of the laser are required. Firstly, we employ a homemade Fabry-Perot (FP) plano-concave cavity to reduce the linewidth of the fundamental laser. The reflectivity of the input plain mirror is 99.3\% and that of the output concave mirror is 99.9\%, resulting in a theoretical finesse of 780. The curvature radius of the concave mirror is 500 mm. The two cavity mirrors are placed 100 mm apart corresponding to a free spectral range of 1.5 GHz. The full width at half maximum (FWHM) of the transmission peak is measured to be 2.35 MHz, corresponding to an experimental finesse of 637. The deviation between the experimental and theoretical results is mainly due to the mode mismatch between the incident beam and the cavity eigenmode. To suppress the thermal and vibration noises, the cavity spacer is made of invar alloy and placed on a vibration-isolating optical breadboard. The fiber-coupled fundamental laser of about 900 $\mu$W is phase-modulated by an electro-optic modulator (EOM) operating at 50 MHz and then coupled into the cavity. Frequency locking to the cavity is based on the well-established Pound-Drever-Hall (PDH) method~\cite{drever1983laser}, with the obtained error signal being shown in Fig.~\ref{Fig2}(a). The peak-to-peak value of the error signal is 464 mV, corresponding to about 1.3 MHz in frequency. After laser frequency being locked to the cavity, the root of mean square (RMS) value of the residual noise is about 10.6 mV, corresponding to a reduced fundamental laser linewidth of about 30 kHz on the order of tens of milliseconds.

Secondly, the resonant frequency of the FP cavity drifts a lot due to various kinds of noises, such as acoustic noise, vibration noise, and thermal noise, etc. Therefore, it is necessary to lock the FP cavity to the UV transition of $^{6}$Li, achieving both linewidth narrowing and frequency stabilization of the UV laser. Due to the relatively small optical cross section of the UV transition of $^{6}$Li, the lithium vapor pipe has to be heated to more than 400 $^{\circ}$C to provide a sufficient optical density for the saturated absorption spectroscopy, generating two severe problems. First, the viewports at both ends of the vapor pipe suffer from the lithium coating problem, leading to a significant drop of the light transmissivity. Second, the consumption of lithium is very high at this temperature, resulting in a short lifetime of the vapor pipe. To solve these problems, we build a 90 cm long vapor pipe with a small inner diameter of 9 mm, reducing the solid angle of the lithium atomic beam. We place a 400 mm long stainless steel mesh inside the vapor pipe to enhance the surface tension of lithium so that the condensed lithium atoms will return back to the hottest region due to the capillary effect. Furthermore, argon is used as buffer gas to prevent the lithium atoms from reaching the viewports. However, there's a tradeoff between the pressure broadening effect and the lithium coating problem. We measure the saturated absorption spectrum with different buffer gas pressures and find a suitable value of $10^{-3}$ mbar. With all of these efforts, the achieved lifetime of the vapor pipe is about half a year, which is acceptable for our experiment.

About 4.5 mW 323 nm laser is used for the UV frequency modulation spectroscopy. The laser passes through an acousto-optical modulator (AOM1: -111 MHz) to shift the laser frequency during the UV MOT process. Then the beam is collimated by a telescope that minimizes the pointing variation during the frequency shifting. A high-Q resonant EOM (Qubig: EO-F20B3-UV) operating at 20.4 MHz is applied for phase modulation. Then, the laser is divided into a pump beam and a probe beam with equal intensities. The two beams counter-propagate through the vapor pipe and the probe beam is detected by a fast photodiode (PD2, EOT ET-2030A). The observed error signal of frequency modulation spectroscopy of the $|2S_{1/2},\ F=3/2\rangle\to |3P_{3/2},\ F'=5/2\rangle$ transition (shown in Fig.~\ref{Fig2}(b)) is used for the cavity length locking. It has a peak-to-peak value of about 380 mV, corresponding to about 12.3 MHz in frequency. After two-stage frequency locking, the RMS value of the residual noise is about 2.1 mV, resulting in a 68 kHz UV laser linewidth on the order of tens of milliseconds.

\begin{figure}[htbp]
\centering
\includegraphics[width=0.8\columnwidth]{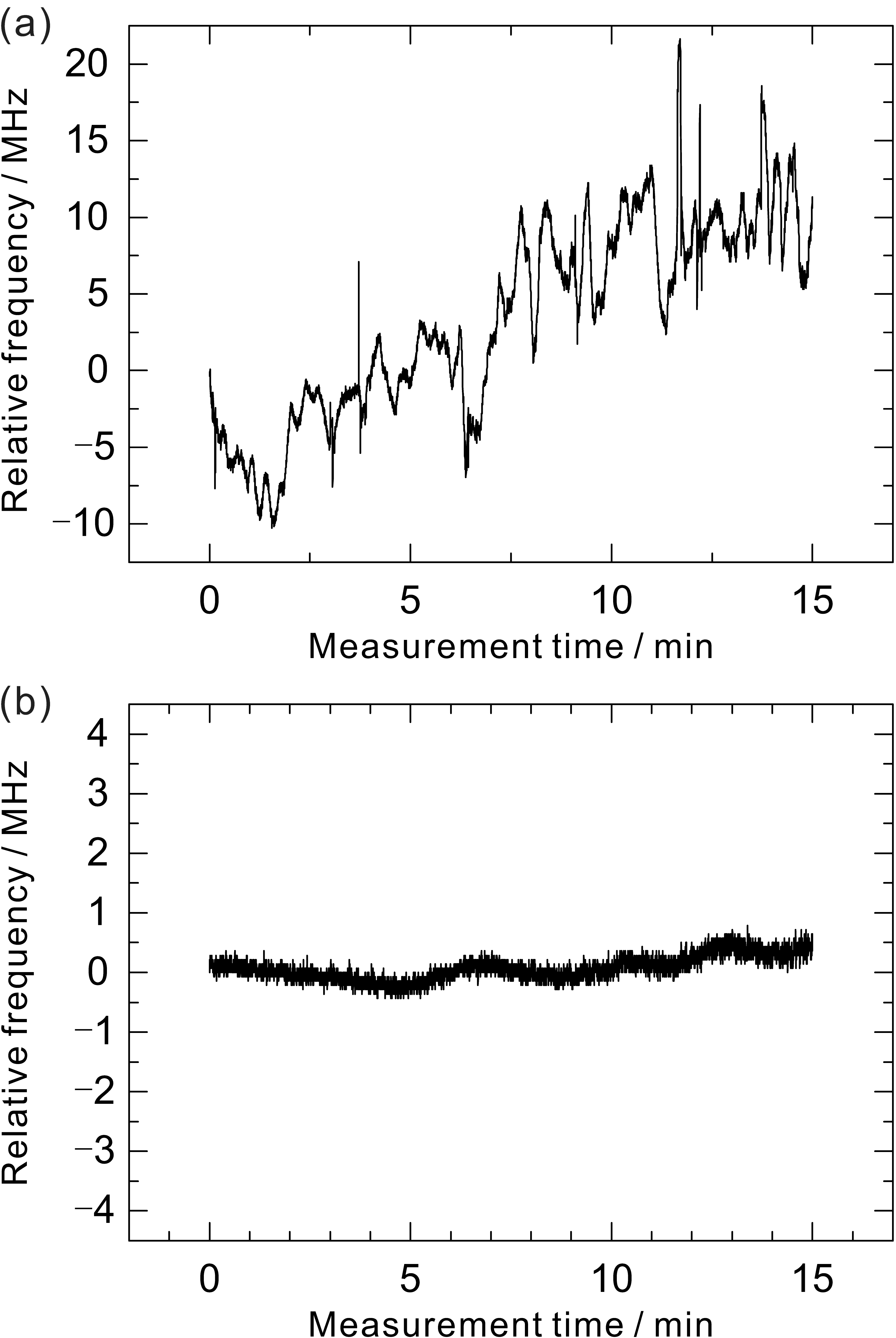}
\caption{(a) shows the long-term drift of the fundamental laser in the free running case. The laser frequency drifts about 30 MHz. (b) shows the long-term drift of the fundamental laser in the frequency locking case which is less than 1 MHz. The measurement duration is about 15 minutes. }
\label{Fig3}
\end{figure}

To characterize the locking performance, we take two additional measurements of the central frequency drift and the laser linewidth. It is known that the error signal obtained by frequency modulation spectroscopy will suffer a Doppler background variation which induces an additional uncertainty in the absolute locking frequency. We use a wavemeter (HighFinesse: WSU-10, measurement resolution: 1 MHz) to determine the long-term drift (see Fig.~\ref{Fig3}). In the free running case, the central frequency drifts about 30 MHz in 15 minutes. When the laser frequency is stabilized, the frequency fluctuation is less than 1 MHz over 15 minutes, within the resolution of the measurement device.

\begin{figure*}[htbp]
\centering
\includegraphics[width=0.8\textwidth]{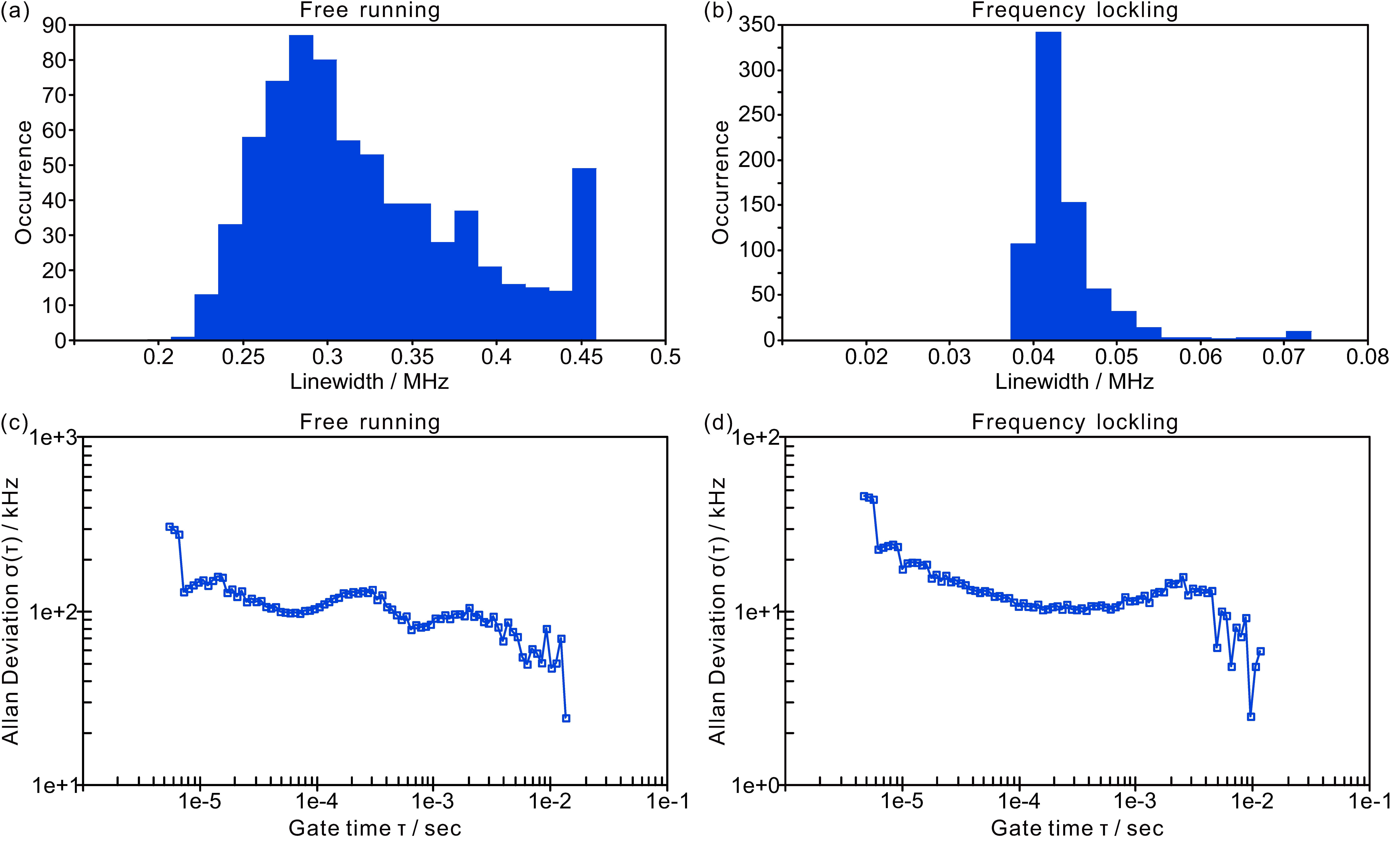}
\caption{(Color online). (a) and (b) show the linewidth measurements of the 646 nm fundamental laser in the free running case and the frequency locking case, respectively. The integration time of each measurement is 100 ms. The linewidth is $333\pm70$ kHz and $46\pm15$ kHz averaged by more than seven hundred times in each case. (c) and (d) show the Allan deviation analysis of the 646 nm fundamental laser in each case.}
\label{Fig4}
\end{figure*}

We further measure the linewidth of the fundamental 646 nm laser with an optical spectrum analyzer (Sirah Lasertechnik: EagleEye). The EagleEye is able to measure linewidths down to 20 kHz. We continuously measure the laser linewidth for more than seven hundred times both in the free running and the frequency locking case, shown in Fig.~\ref{Fig4}(a) and (b). The integration time is 100 ms for each measurement. The measured laser linewidth is $333\pm70$ kHz in the free running case and $46\pm15$ kHz in the frequency locking case, respectively. The linewidth of the UV laser is approximately two times as large as that of the fundamental laser which agrees with our previous results. Moreover, we analyse the Allan deviation in each case which are shown in Fig.~\ref{Fig4}(c) and (d). The results indicate that the noises are strongly suppressed by the servo loop.

The main SHG output is split into two parts by a half-wave plate (HWP) and a polarization-beam-splitter (PBS). The cooling laser passes through AOM2 to shift the frequency by -114 MHz while the repumping laser passes through AOM3 to shift the frequency by +114 MHz. Afterwards, the two laser beams are superimposed on a PBS. An aspheric lens and a 25 $\mu$m precision pinhole is used to filter out the higher-order transverse modes of the beam, resulting in a TM00 mode laser beam. Then the laser beam is collimated by another lens, yielding a 1/e$^{2}$ radius of 4.7 mm. Finally, the UV laser is divided into six balanced laser beams by a series of PBSs and HWPs. The resulting six UV MOT laser beams are superimposed with the red MOT laser beams through six custom-designed dichroic mirrors, where 323 nm lasers are reflected while those of 671 nm, 767 nm, and 770 nm are transmitted. To minimize the loss of laser power, all optics for the UV MOT are custom-designed for high transmission rates and damage thresholds.

\section{Two-stage MOT \& magnetic transport}

\begin{figure}[htbp]
\centering
\includegraphics[width=\columnwidth]{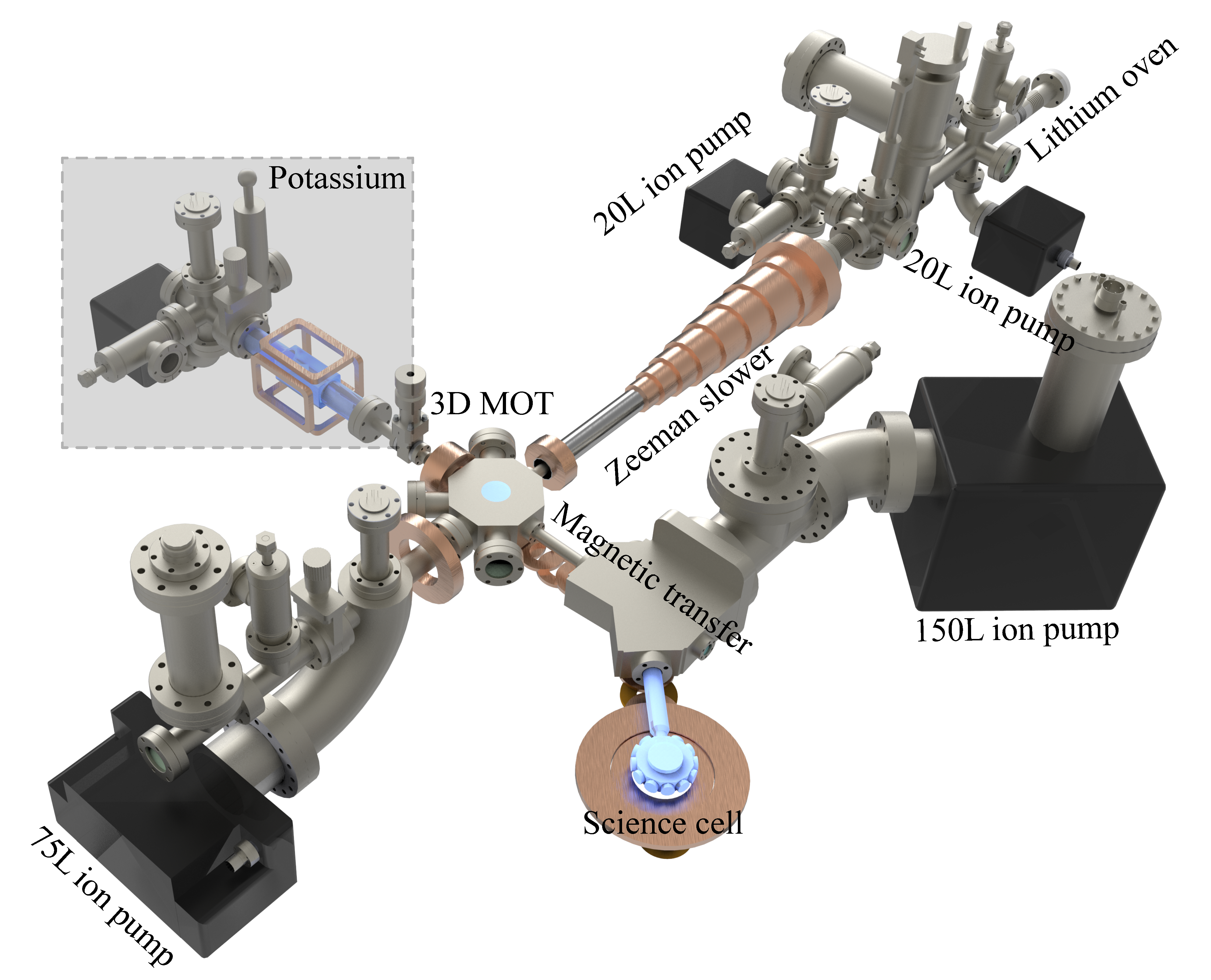}
\caption{(Color online). Simplified layout of the vacuum assembly. The vacuum setup for lithium consists of an oven, a Zeeman slower, a 3D MOT chamber, a magnetic transport chamber, and a full glass science cell. The shadow region is for potassium experiments.}
\label{Fig5}
\end{figure}

\begin{figure}[htbp]
\centering
\includegraphics[width=\columnwidth]{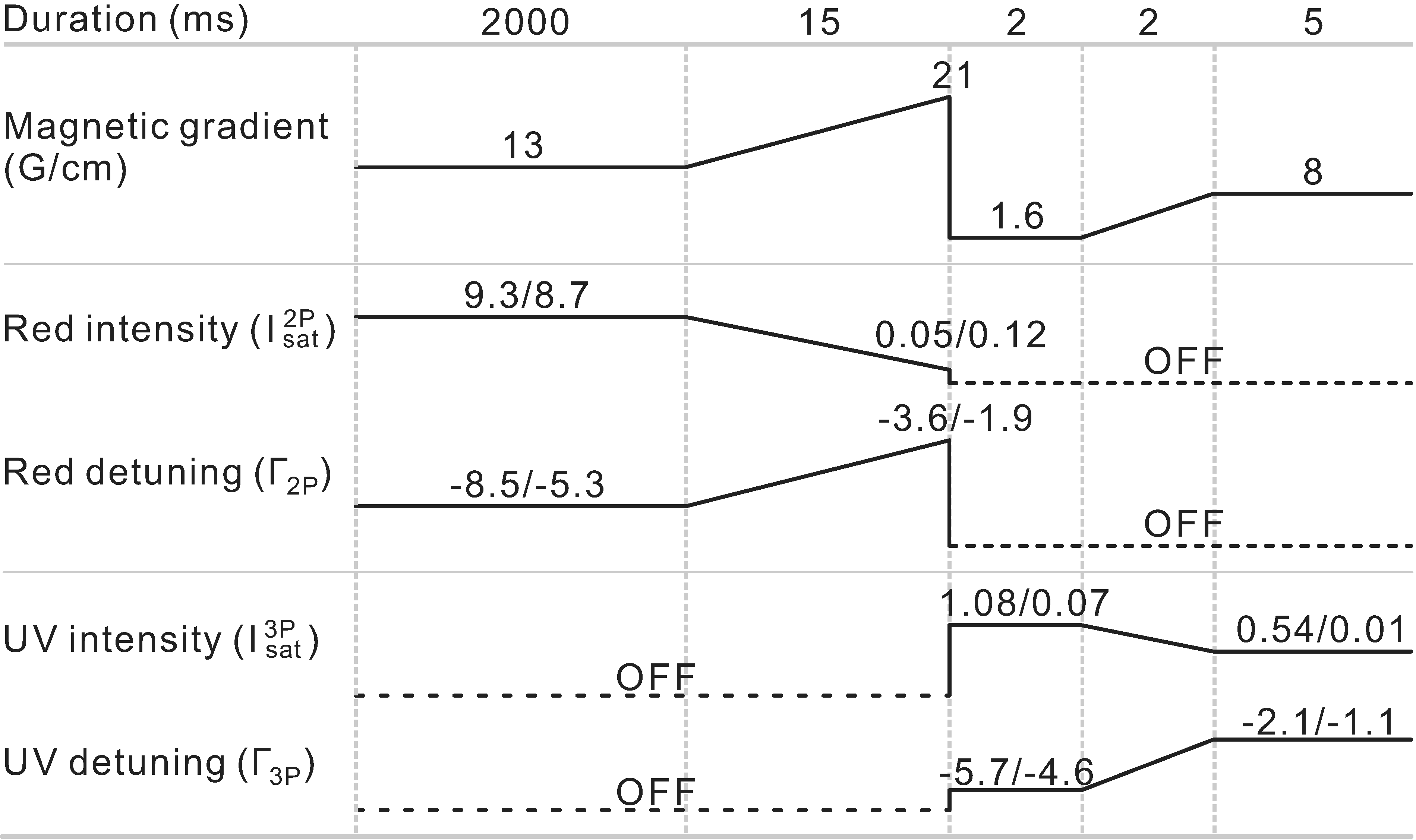}
\caption{Experimental sequence and parameters of the laser cooling phase. The intensities and detunings of the cooling and repumping lasers are specified, respectively.}
\label{Fig6}
\end{figure}

The vacuum layout is shown in Fig.~\ref{Fig5}. The experimental sequence starts from generating an intense slow lithium beam through a spin-flip Zeeman slower. In order to match the divergence of the slowing laser with that of the lithium beam, we focus the slowing laser using a lens pair. The 1/e$^{2}$ radius is about 8 mm at the center of the MOT chamber and the focal point is 80 cm away from the MOT center where the Zeeman slower begins to take effect. The peak intensities of the cooling and repumping lasers are 15.7 $I_{sat}^{2P}$ and 5.5 $I_{sat}^{2P}$, respectively, where $I_{sat}^{2P}$=2.54 mW/cm$^{2}$ is the saturation intensity of the D2 transition of $^{6}$Li. Both the cooling and repumping lasers are $\sigma^+$ polarized and the optimal detuning is $-80\ \Gamma$, where $\Gamma=2\pi \times 5.87$ MHz is the D2 transition natural linewidth of $^6$Li. The Zeeman-slower viewport is heated to 230 $^{\circ}$C to prevent the inner surface from being coated by the lithium flux. The optimized lithium flux is more than $2\times10^{10}$ atoms per second when the oven is heated to 450 $^{\circ}$C. The operating temperature is less than 400 $^{\circ}$C in order to prolong the oven's lifetime. Then, a conventional D2-line red MOT is applied to capture the atomic cloud from the cold lithium flux. The total power of the MOT laser is about 370 mW. By using a series of HWPs and PBSs, the MOT laser is divided into six independent laser beams with a 1/e$^{2}$ radius of 9 mm. The intensities of cooling and repumping components are approximately identical.  After a 2 seconds loading phase, we collect about $2.4\times10^{9}$ atoms at a temperature of 1.5 mK and a peak density of $1.2\times10^{10}$ cm$^{-3}$. To further increase the phase space density (PSD), we employ a compressed-MOT (CMOT) in the following 15 ms. During this phase, the magnetic gradient is ramped up from 13 G/cm to 21 G/cm. Moreover, we decrease the laser intensity and the detuning to suppress light-assisted collisions (see Fig.~\ref{Fig6} for details). This step yields a twofold increase in the peak density while the temperature decreases to 252 $\mu$K. The calculated PSD is about $2.2\times10^{-6}$.

The time sequence for the UV MOT contains three steps, i.e., loading, compressing, and cooling (see Fig.~\ref{Fig6}). At the end of the CMOT phase, we shut off the red MOT laser beams, switch on the UV MOT laser beams, and reduce the magnetic field gradient simultaneously.  Due to the optical losses of AOMs and other optical elements, the total laser power acting on the atoms is about 33 mW for the UV MOT. For each beam, the peak intensity of the cooling (repumping) component is 1.08 $I_{sat}^{3P}$ (0.07 $I_{sat}^{3P}$), where $I_{sat}^{3P}=13.8$ mW/cm$^{2}$ is the saturation intensity of the UV transition. During the 2 ms loading phase, the maximum intensity and largest detuning of the cooling (repumping) laser are used to capture as many atoms as possible from the CMOT. The atom number in the UV MOT is about $1.3\times10^{9}$, corresponding to a capture efficiency of 72\%. In the compressing phase, the magnetic field gradient is increased from 1.6 G/cm to 8 G/cm in 2 ms. The cooling (repumping) intensity of each beam is linearly decreased from 1.08 $I_{sat}^{3P}$ (0.07 $I_{sat}^{3P}$) to 0.54 $I_{sat}^{3P}$ (0.01 $I_{sat}^{3P}$) while the cooling (repumping) detuning is reduced from $-5.70\ \Gamma_{3P}$ ($-4.70\ \Gamma_{3P}$) to $-2.10\ \Gamma_{3P}$ ($-1.10\ \Gamma_{3P}$) simultaneously. The compressing phase is essential for increasing the peak density of the lithium cloud. Further UV MOT cooling is achieved by holding the intensity, detuning, and magnetic field gradient for another 5 ms.

With time-of-flight measurements, the obtained temperature of the UV MOT is about T=58 $\mu$K and the peak density is about $5.5\times10^{10}$ cm$^{-3}$, resulting in a PSD of $4.5\times10^{-5}$ which is twentyfold larger than that of the CMOT. We also find that by further decreasing the detuning of the UV laser during the cooling phase, a much lower temperature of 34 $\mu$K can be attained at the cost of atom number loss. In this case, the attained atom number is $3.5\times10^{8}$ and the PSD decreases to $1.1\times10^{-5}$. On the other hand, we can capture more than $1.6\times10^{9}$ atoms in the UV MOT by using a larger laser detuning during the loading phase. However, the temperature increases to 84 $\mu$K which results in a lower PSD. The parameters and time sequence depicted in Fig.~\ref{Fig6} are optimized for the following D1 optical pumping and magnetic trapping phase.

At the end of the UV MOT, the atoms are randomly distributed in the hyperfine ground state. To load the atoms into the magnetic trap, we employ a high field D1 optical pumping method to prepare all the atoms in the stretched state $|F=3/2,\ m_{F}=3/2\rangle$. The applied atomic transitions, $|2S_{1/2},\ F=3/2\rangle\to |2P_{1/2},\ F'=3/2\rangle$ and $|2S_{1/2},\ F=1/2\rangle\to |2P_{1/2},\ F'\\=3/2\rangle$, serve as D1 pumping and repumping frequencies, respectively. Experimentally, we suddenly turn on a homogeneous magnetic field of 15 G and shine a pair of balanced $\sigma^+$ polarized counter-propagating D1 laser beams on the atoms over 50 $\mu$s. Then the magnetic quadrupole trap is quickly ramped to a gradient of 100 G/cm in 20 ms. With the D1 optical pumping, we can transfer about 90\% lithium atoms from the UV MOT to the magnetic trap. After being confined in the magnetic trap, the lithium cloud is magnetically transported from the MOT chamber to the science cell that has a good vacuum environment and large optical access.

\begin{figure*}[htbp]
\centering
\includegraphics[width=0.8\textwidth]{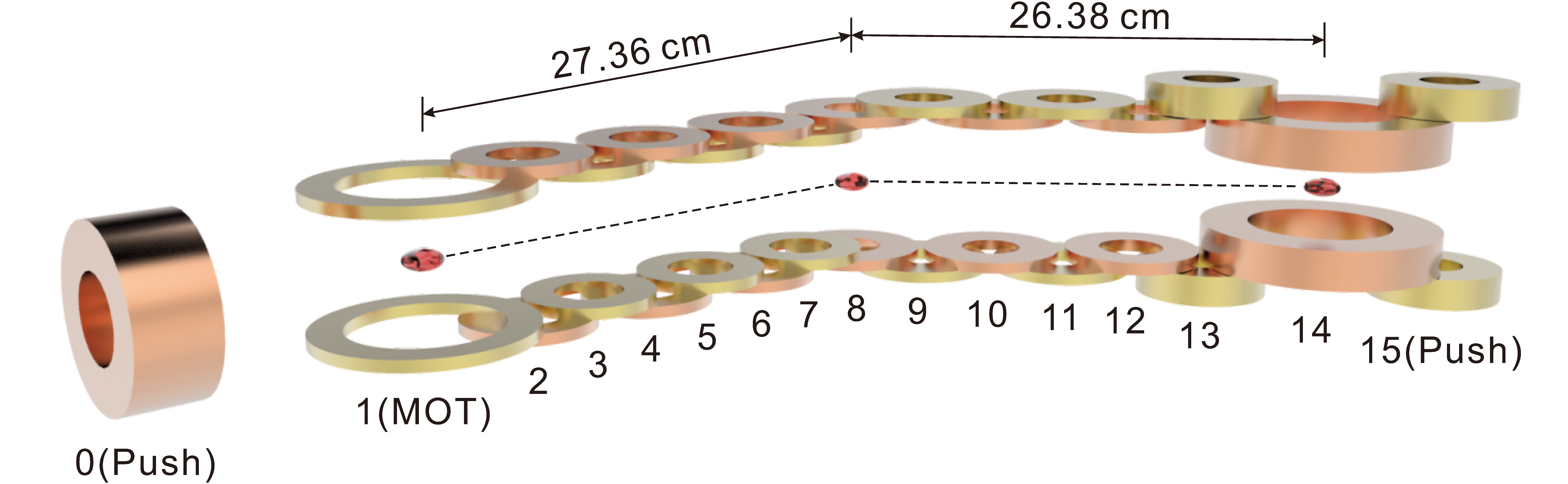}
\caption{(Color online). Simplified layout of the magnetic transport system. The coil pairs are labelled from No. 0 to No. 15. The dashed line represents the transport path.}
\label{Fig7}
\end{figure*}

The magnetic transport system consists of sixteen pairs of overlapping coils which provides a moving magnetic trapping potential (see Fig.~\ref{Fig7})~\cite{greiner2001magnetic}. The atomic clouds are transported from the center of the MOT (No. 1) coil pair to that of the No. 14 coil pair. Two transport arms form an angle of $135^{\circ}$ with a total distance of 53.74 cm. The pushing coils (No. 0 and No. 15) are used to create an additional magnetic gradient field along the transport direction, which are necessary to make the aspect ratio of the atomic cloud varying smoothly in the beginning and ending of the transport process. We optimize the current running in the each coil pairs, total transport time and magnetic field gradient to achieve the best efficiency. Finally, we can transport $8.1\times10^8$ lithium atoms to the science cell in a 3 s transporting process, corresponding to an transport efficiency of 70.3\%. The major loss of atoms is due to the small diameter (5 mm) differential pumping tube in the transport path, where hot atoms are blocked by the wall of the tube. The measured temperature is about 296 $\mu$K at a 198 G/cm magnetic field gradient in the science chamber, serving as an excellent starting point for further evaporative cooling. Furthermore, we also transport the lithium atoms without the implementation of narrow-linewidth cooling. The achieved transport efficiency is limited to 23.6\%, showing the advantage of the narrow-linewidth cooling.

\section{Conclusion}

In summary, we combine a red MOT with a UV MOT of $^{6}$Li using the $2S_{1/2}\to 3P_{3/2}$ transition at 323 nm. The narrow-linewidth cooling allows us to achieve a sub-Doppler temperature of the $^{6}$Li with respect to the D2 transition. The obtained atom number of the UV MOT is about $1.3\times10^{9}$ with a temperature of 58 $\mu$K, demonstrating a twentyfold enhancement of the PSD. Furthermore, the lithium atoms are magnetically transported to a science chamber with a good vacuum environment and large optical access that facilitate future studies of lithium gases, such as optical lattices~\cite{bloch2005ultracold} and quantum gas microscopy~\cite{greif2016site}. Thanks to the low temperature and high PSD of $^{6}$Li atoms, a threefold increase in transport efficiency is achieved. The obtained $8.1\times10^8$ atoms with a temperature of 296 $\mu$K at a 198 G/cm magnetic field gradient in the science chamber is an excellent starting point for the production of large $^{6}$Li quantum degenerate gases. In the near future, we will combine the UV MOT of $^{6}$Li with the gray molasses of $^{41}$K to achieve an even larger two-species Bose-Fermi superfluid~\cite{yao2016observation}.

\begin{acknowledgements}
This work has been supported by the National Natural Science Foundation of China, the Chinese Academy of Sciences, and the National Fundamental Research Program (under Grant No. 2013CB922001). X.-C.Y. acknowledges support from the Alexander von Humboldt-Stiftung.
\end{acknowledgements}

\end{document}